\documentclass[11pt,twoside]{article} 
\usepackage{asp2004}
\usepackage{epsf}
\usepackage{psfig}
\usepackage{lscape} 

\begin{document} 
\title{3D-Kinematics of White Dwarfs from the SPY-Project}
\author{E-M. Pauli, U. Heber} 
\affil{Dr. Remeis-Sternwarte, Astronomisches Institut
der Universit\"at Erlangen-N\"urnberg, Sternwartstra\ss e 7,
D-96049 Bamberg, Germany}
\author{R. Napiwotzki} 
\affil{Department of Physics \& Astronomy, University of Leicester, 
University Road, Leicester LE1 7RH, UK }
\author{M. Altmann} 
\affil{Departamento de Astronomia, Universidad de Chile, Camino El
Observatorio 1515, Las Condes, Chile}
\author{M. Odenkirchen} 
\affil{Max-Planck-Institut f\"ur Astronomie, K\"onigstuhl 17, 69117 Heidelberg,
 Germany} 
\begin{abstract} 
We present kinematics of a sample of $398$ DA
white dwarfs from the SPY project (ESO SN Ia Progenitor surveY)
and discuss kinematic criteria for a distinction of thin disk,
thick disk, and halo populations.
This is the largest homogeneous sample of white dwarfs for which 3D
space motions have been determined.
Radial velocities and spectroscopic distances obtained by the
SPY project are combined with our measurements of proper motions
to derive 3D space motions.
Galactic orbits and further kinematic parameters are computed.
Our kinematic criteria for assigning population membership are
deduced from a sample of F and G stars taken from the literature
for which chemical criteria can be used to distinguish
between thin disk, thick disk and halo.
Our kinematic population classification scheme is based on the
position in the $U$\/-$V$-velocity diagram, the position
in the $J_{\mathrm{z}}$-eccentricity diagram and the Galactic orbit.
We combine this with age estimates and find seven halo and
$23$ thick disk white dwarfs.
\end{abstract}

\section{Introduction}

White dwarfs are presumably valuable tools
in studies of old populations such as the halo and the thick disk,
since the fraction of old stars among white dwarfs is higher than
among main-sequence stars. An open issue is the fraction of white dwarfs in the
thick disk and halo populations and their
percentage of the total mass of the Galaxy.
Studies of their kinematics can help to determine the fraction
of the total mass of our Galaxy contained in
the form of thick disk and halo white dwarfs which allows 
the role of white dwarfs in the dark matter problem to be studied.


The common problem of kinematic studies of white dwarfs is the lack
of radial velocity measurements. Especially deviating conclusions derived
from the white dwarfs of the Oppenheimer et al. (2001) sample
demonstrate that different assumptions on the values of $v_{\rm rad}$
can produce different fractions of halo and thick disk stars
and thus have a strong impact on the determination
of the white dwarf halo density.
Therefore a sample of white dwarfs with known radial velocity measurements
is needed in
order to obtain the full 3D kinematic information. 
Pauli et al. (2003) presented a complete 3D kinematical study of 107 white 
dwarfs. Here we extend this investigation to about 400 stars.  

\section{3D kinematics and population classification}
The ESO Supernova Ia Progenitor survey (SPY, Napiwotzki et al., 2003) 
secured high resolution spectra
of more than 1000 degenerate stars in order to test the double degenerate
scenario for the SN Ia progenitors. The sharp NLTE absorption core of the 
H$\alpha$ line allowed accurate radial velocities to be measured (Napiwotzki
et al., in prep.).
A spectroscopic analysis yielded atmospheric parameters, masses and
gravitational redshift (see Koester et al., 2001).
Proper motions for $202$ stars have been measured
whereas the rest was extracted from catalogs (USNO--B, UCAC2,
the SuperCOSMOS Sky Survey, 
the Yale Southern Proper Motion, 
the revised NLTT and LHS catalogs).
We calculate individual errors of kinematic parameters by means of
Monte Carlo error propagation.

Pauli et al. (2003) presented a new 
population classification scheme based on the $U$\/-$V$-velocity diagram,
the $J_Z$-eccentricity-diagram and the Galactic orbit.
For the computation of orbits and kinematic parameters we used the code
by Odenkirchen \& Brosche (1992) based on the Galactic potential of 
Allen \& Santillian (1991). 
The classification scheme is based on a calibration sample of
main-sequence stars.

Unlike for main-sequence stars the population membership of white dwarfs
can not be determined from spectroscopically measured metalicities.
Therefore we have to rely on kinematic criteria, which have to be calibrated 
using a suitable calibration sample of
main-sequence stars.
In our case this sample consists of $291$ F and G main-sequence stars from
Edvardsson et al. (1993) and Fuhrmann (2004, and references cited therein)

\begin{figure}
\vspace{17.5cm}
\includegraphics{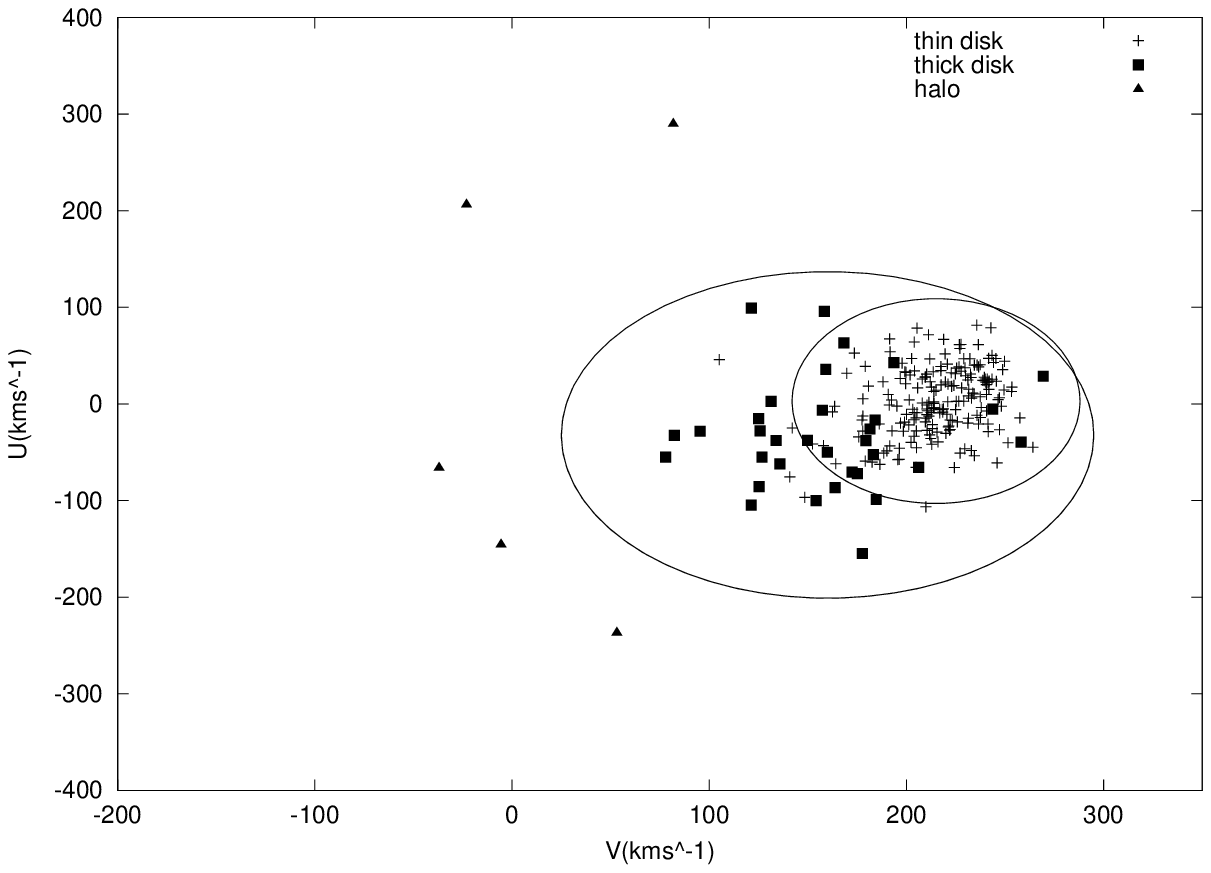}
\includegraphics{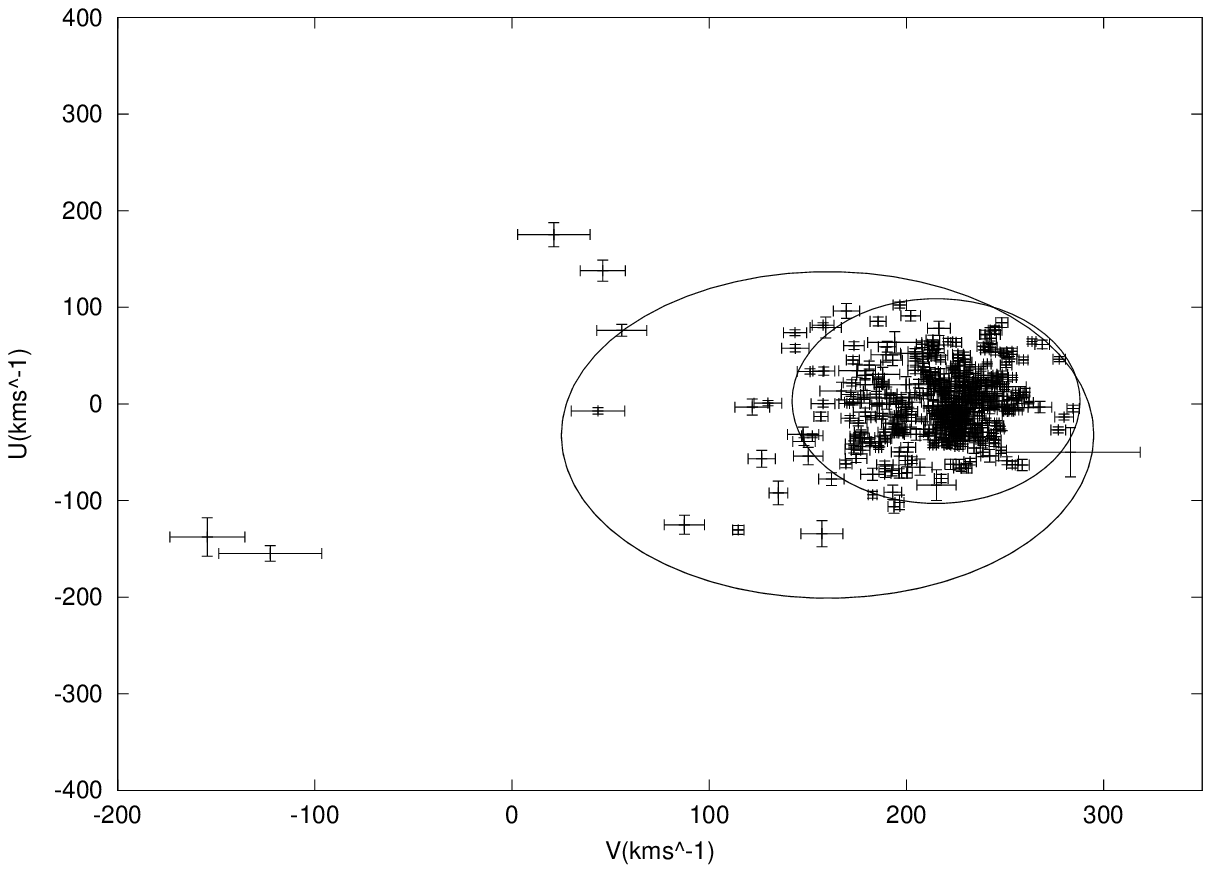}
\caption{$U$\/-$V$-velocity diagram for the main-sequence stars (top) and
for the white dwarfs (bottom);
	dashed lines: $3\sigma_{\rm thin}$-, $3\sigma_{\rm thick}$-contours.
}
	\label{uvms}
\end{figure}

Halo and thick disk stars can be
separated by means of their [Fe/H] abundances,
they possess a higher [Mg/Fe] ratio than thin disk stars (see Pauli et al.,
2003 for details).
In Fig.~\ref{uvms} (top panel) $U$ is plotted
versus $V$ for the main-sequence stars.
For the thin disk and the thick disk stars the mean values and standard
deviations (3$\sigma$) of the two velocity components have been calculated.
The values for the thin disk are:
$\left<U_{\rm ms}\right>=3~\pm 35~\rm{km~s^{-1}}$,
$\left<V_{\rm ms}\right>=215~\pm 24~\rm{km~s^{-1}}$,
The corresponding values for the thick disk are:
$\left<U_{\rm ms}\right>=-32~\pm 56~\rm{km~s^{-1}}$,
$\left<V_{\rm ms}\right>=160~\pm 45~\rm{km~s^{-1}}$,
Indeed, nearly all thin disk stars stay inside the
$3\sigma_{\rm thin}$-limit and all halo stars lie outside
the $3\sigma_{\rm thick}$-limit, as can be seen from Figure~\ref{uvms}.
In the $J_Z$-eccentricity-diagram three regions (A, B or C) can be defined
(not shown here, but see Pauli et al. 2003)
which host thin disk, thick disk and halo stars, respectively. 
The Galactic orbits of thin and thick disk stars and halo stars 
differ in a characteristic way allowing another classification criterion to be
defined.

The halo candidates are all white dwarfs that are either
situated outside the $3\sigma$-limit of the thick disk in the
$U$\/-$V$-velocity diagram or that lie in Region~C in
the $J_Z$-eccentricity diagram and have halo type orbits.
Seven white dwarfs fulfill these conditions and are therefore assigned to the 
halo.
Two white dwarfs
are on retrograde orbits characterized by a negative value
of $V$ and $J_{\rm Z}$. 

Thick disk white dwarfs lie either
outside the $3\sigma$-limit of the thin disk in the
$U$\/-$V$-velocity diagram or lie in Region~B in
the $J_Z$-eccentricity diagram and have thick disk type Galactic orbits.
Twenty-seven of them are classified as thick disk members. 
All the remaining are assumed to belong to the thin disk,
leaving us with seven halo, $27$ thick disk out of the $398$ SPY white dwarfs.

Ages are another criterion for population membership.
Halo and thick disk white dwarfs need to be old stars. Since the cooling
ages of the of most white dwarfs in the SPY sample are small ($<$ 10$^9$yrs)
they have evolved from long-lived, i.e. low mass stars and hence must
themselves be of low mass. All halo and 23 of the 27 thick disk
white dwarfs have sufficiently low masses and are therefore old. 
Hence we assign only these $23$ white dwarfs to the thick disk, the others 
are either kinematically misclassified thin disk stars or runaway
stars.
This leaves us with a fraction of $1.8\%$ halo and
$5.8\%$ thick disk white dwarfs.

\section{Discussion}

We have applied the classification scheme
developed in Pauli et al. (2003) to kinematically
analyze a sample of about $400$ DA white dwarfs from the SPY project.
Combining the three kinematic criteria position in the $U$\/-$V$-diagram,
position in the $J_Z$-$e$-diagram and Galactic orbit with age estimates
we have found seven halo and $23$ thick disk members.

The velocity dispersions that for the thin disk
white dwarfs are well compatible with the ones of Soubiran et al. (2003).
The same is the case for the asymmetric drift and the velocity
dispersions of the thick disk.
For the halo white dwarfs $\sigma_{\rm U}$ and $\sigma_{\rm V}$ 
are similar to the values derived by Chiba \& Beers (2000) for halo stars while
our $\sigma_{\rm W}$ is much smaller, probably due to low number statistics.
Hence the kinematic parameters
of the white dwarfs of the three different populations
do not differ much from main-sequence samples.

\acknowledgements{M.A. is financially supported by FONDAP~1501~0003.}

\end{document}